\documentclass[preprint]{elsarticle}%
\usepackage{graphicx}
\usepackage{bm}
\usepackage{epsf}
\usepackage{rotating}
\usepackage{epsfig,graphics,rotate,color}
\usepackage{amssymb}
\usepackage{amsmath}
\usepackage{amsfonts}
\usepackage{array,hhline,dcolumn}%
\providecommand{\U}[1]{\protect\rule{.1in}{.1in}}

\begin{document}
\title{Demonstration of a Lightguide Detector for Liquid Argon TPCs}

\author{L. Bugel,  J.M. Conrad, C. Ignarra, B.J.P. Jones, T. Katori, T.Smidt and H.-K. Tanaka\footnote{Now at Brookhaven National Laboratory}}
\address{Physics Dept., Massachusetts Institute of Technology,
Cambridge, MA 02139}

\begin{abstract}
  We report demonstration of light detection in liquid argon
  using an acrylic lightguide detector system.  This opens the
  opportunity for development of an inexpensive, large-area light
  collection system for large liquid argon time projection chambers.
  The guides are constructed of acrylic, with TPB embedded in a
  surface coating with a matching index of refraction.  We study the
  response to early scintillation light produced by a 5.3 MeV
  $\alpha$.  We measure coating responses from 7 to 8 PE on average,
  compared to an ideal expectation of 10 PE on average.  We estimate
  the attenuation length of light along the lightguide bar to be
  greater than 0.5 m.  The coating response and the attenuation length
  can be improved; we show, however, that these results are already
  sufficient for triggering in a large detector.

\end{abstract}
\maketitle

\section{Introduction}

Scintillation photons are copiously produced when a charged particle
traverses liquid argon.   This occurs through two processes \cite{Martin, Molchanov, Recomb}.
An argon atom may be
excited and then form a dimer which radiatively decays,
\begin{equation}
{\rm Ar}^* + {\rm Ar} \rightarrow {\rm Ar}_2^*
  \rightarrow 2 {\rm Ar} + \gamma.
\end{equation}
Alternatively,  an argon atom may be ionized, then form an ionized molecule, which subsequently
picks up an electron through recombination, forming a dimer which radiatively
decays,
\begin{equation}
{\rm Ar}^+ + {\rm Ar} \rightarrow {\rm Ar}_2^+ + e^- 
\rightarrow {\rm Ar}_2^* \rightarrow 2 {\rm Ar} + \gamma.
\end{equation}
Singlet and triplet states of Ar$_2^*$ will be formed with with
emission lifetimes of 6~ns and 1.6 $\mu$s, respectively.  
The emitted photon has wavelength in the vacuum ultraviolet,
at 128 nm \cite{ICARUS}.

In an ultra-large liquid argon time projection chamber (LArTPC), the
early scintillation light can be used as a trigger for neutrino events
or as a veto for cosmic rays.  The light must be shifted from the
vacuum ultraviolet to the visible in order to be detected.  The most
common LArTPC designs use 8 inch or larger cryogenic phototubes, which
collect light that has been shifted to the visible by
1,1,4,4-Tetraphenyl-1,3-butadiene (TPB) \cite{ICARUS, MicroBooNE}.

In this paper, we present a new, cost-effective alternative for light
collection in LArTPCs. This uses lightguides which are made from
extruded acrylic covered with a polystyrene skin embedded with TPB.
Multiple acrylic bars can be bent to guide light adiabatically to
a single 2 inch cryogenic phototube (PMT).

Lightguides have a thinner profile than the PMT systems, occupying
less space in an an LAr vessel and potentially allowing more fiducial
volume.  This is important in the case of very large detectors
proposed for long baseline neutrino experiments \cite{Barger}, where
fiducial volume is at a premium.

In the discussion below, we describe the design of the lightguides and
the teststand.  We describe the conversion from measured waveform
pulseheight to photoelectrons (PE).  We predict the ``ideal'' average
signal achievable with the present design.  We compare this to the
measured value. We consider the losses due to transport along the bar.
Lastly, we consider how the guides might be used to form paddles for
installation in a large liquid argon detector to provide a
pulseheight-based trigger.

\section{Design of Lightguides and Teststand \label{test}}

\begin{figure}[t]\begin{center}
{\includegraphics[width=3.in]{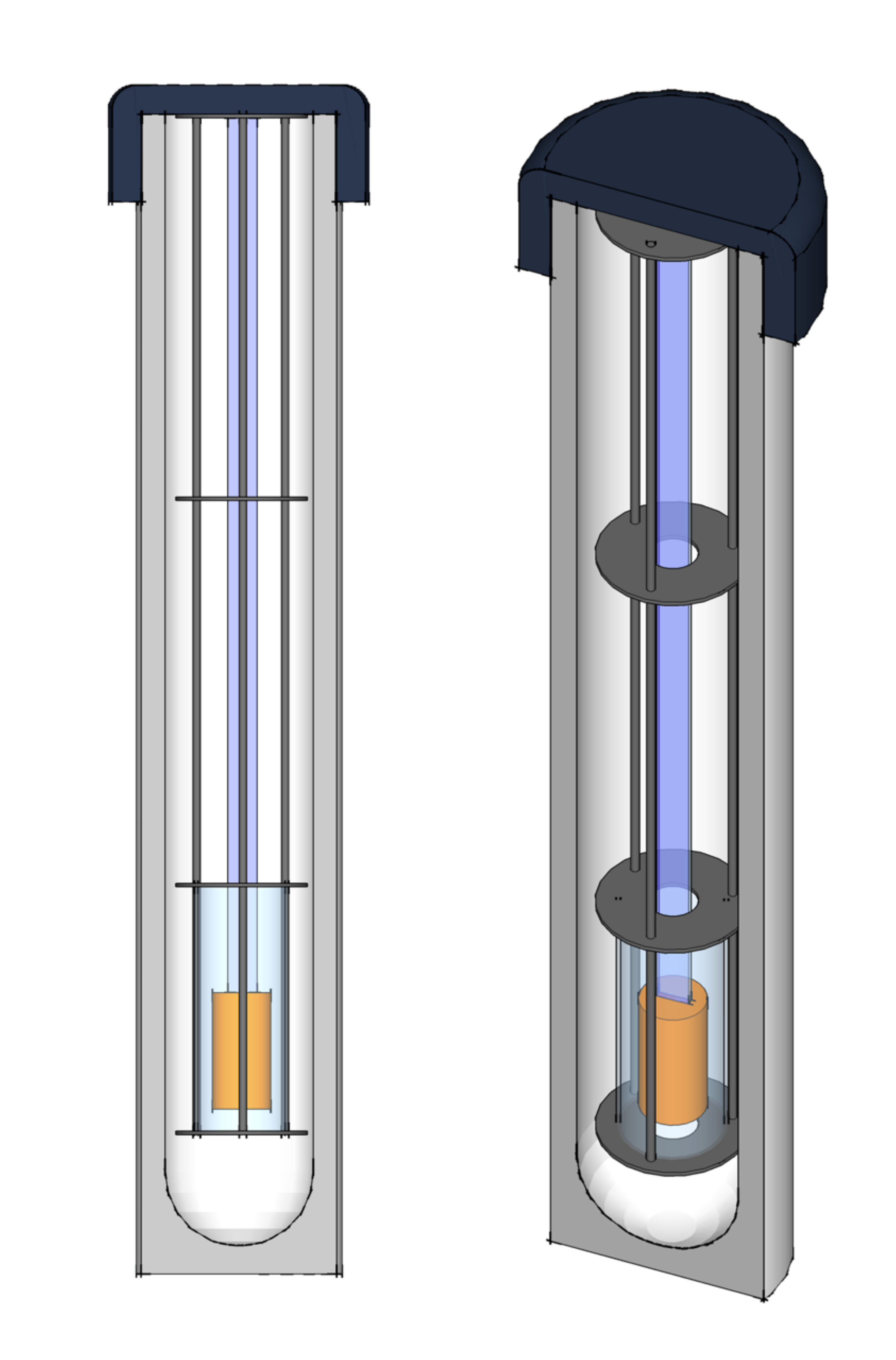}} 
\end{center}
\caption{Two views of the dewar teststand showing the lightguide and PMT.
\label{dewar} }
\end{figure}

To achieve total internal reflection, thereby trapping and guiding the
light, the material in which the light is produced must have a higher
index of refraction than the surrounding medium.  We use acrylic,
which has an index of refraction of $n=1.49$.  Therefore, the TPB must
be embedded in a matrix with a good match in $n$. This is applied as a
smooth skin on the acrylic bars.  We use polystyrene which has an
index of refraction of $n=1.59$.  The bar is immersed in liquid argon,
which has an index of refraction in visible wavelengths of $n=1.23$
\cite{index}.

The lightguides consist of McMaster-Carr extruded rectangular acrylic
bars. For these tests, the dimensions are 0.48 cm (3/16 inch) $\times$
2.54 cm (1 inch) $\times$ 55 cm (21.6 inch).  The acrylic is resilient
to cryogenic temperatures and shows no evidence of crazing. The ends
of the lightguides are polished.

The TPB-embedded skin is applied as a liquid along the length
of the bar.  TPB and polystyrene are
combined in a 1:3 ratio by mass and dissolved in toluene.  The
polystyrene is Dow Styron 663.  The TPB doping is limited to this
level because we have found that more equal ratios of TPB to
polystyrene lead to crystallization of the TPB, which spoils the
optical properties of the skin.  TPB has negligible self-absorption
\cite{berlman}, and so, in principle, guiding can have low
attenuation.

This TPB-polystyrene mixture is sprayed onto the acrylic bar using a
high volume low pressure (HVLP) gravity feed spray gun running on dry
air at 35 psi.  The design in this paper used a single coat.  The bars
are not buffed.

One expects less visible light per UV photon from TPB embedded in a matrix compared to TPB
which is evaporatively applied, because the material of the matrix will
absorb some UV photons.  We have tested the light-response of the coating
using a vacuum spectrometer which illuminates an acrylic plate at 128
nm.  We compare the response of the TPB-polystyrene skin to pure TPB
applied as an evaporative coating.  We find the ratio of light from
our TPB-polystyrene mix to evaporative coating is $\sim 10\%$.  Higher
ratios of TPB to polystyrene are found to give more light, but suffer
from crystalization on the surface.

The light is guided to an R7725-mod Hamamatsu cryogenic PMT.  The test
system is shown in Fig.~\ref{dewar}.  In the liquid argon, which has
density of $1400$ kg/m$^3$, the PMT and the acrylic bar will both
float.  A single bar is held loosely in vertical position with the PMT 
below it in a guide tube.  Because of buoyancy, the PMT presses up against the bar,
which is held in place at the top, naturally making a good 
connection between the glass face of PMT and the acrylic bar.

The PMT is an 7725-mod Hamamatsu, 10-stage phototube \cite{Hamamatsu}
run at 1800~V.  A
cryogenic PMT is required because below 150 K, standard PMTs cease to
produce signal \cite{Meyer}.  This problem is solved in a cryogenic
PMT with a platinum undercoating that allows electron replenishment at
liquid argon temperatures.

The custom-designed base has a dynode chain made only with metal film
resistors and NP0/C0G type capacitors to minimize temperature
dependences. The PC board is made with Roger 4000 series woven glass
reinforced material for lower contamination and copper-like thermal
coefficient of expansion. The circuit elements were tested for
resilience under many cryo-cycles. We use 95$\Omega$ RG180 cable, as
is used in the MicroBooNE experiment. This Teflon-jacketed, 0.14"
diameter cable is typically used in liquid Argon experiments to avoid
contamination.  The cable was successfully tested up to 5kV DC.  The
DC high voltage and the signal are run on this single cable, which
minimizes feed-throughs into a cryogenic system. The signal is picked
off using a blocking capacitor, located outside of the cryogenic
system.

The teststand consists of a glass dewar which is 100 cm tall and 
14 cm inner diameter.  The PMT and bar are lowered into the dewar via a
weighted stand so that the system has negative buoyancy in liquid
argon.  The entire system is in a light-tight box.

The light is produced via 5.3 MeV $\alpha$ particles produced from a
$^{210}$Po source mounted in a plastic disk \cite{Unitednuclear}.  The
total activity of the source was 925 Hz.  However, the geometry of the
source holder reduces the total number of $\alpha$s that can be
observed.  The source is electroplated onto silver foil that is
open-mounted (no window) in the recess, or ``well,'' of the plastic
disk.  The height of the well-wall is 3 mm.  The $\alpha$s which
emerge into the well will traverse only 50$\mu$m in LAr, and so will
not exit the well.  To directly measure the count rate for $\alpha$s
leaving the source, we fit a small piece of solid scintillator into
the well and attach this to a PMT.  The rate was measured to be $359
\pm 11$ Hz.

The disk-source is mounted in a holder which can be moved along the
bar, and which offsets the source from the face of the bar by a total
of 6 mm.  The holder is held in place by two screws which press
against the bar.  The geometry of the source-disk limits the
acceptance for UV light produced in the argon to hit the bar.
However, because of the close proximity between source and bar, if the
UV light leaves the well, the photons will hit the bar.

The $\alpha$ source only produces a measurable rate of signal above
background from the bar when immersed in liquid argon.  We were not
able to measure a rate above background from the source in either liquid
nitrogen or in air.  When running in liquid argon, if the liquid level
is allowed to drop through evaporation, the signal from the $\alpha$
sources stops when the liquid level drops below the source.

\begin{figure}[t]\begin{center}
{\includegraphics[width=2.in]{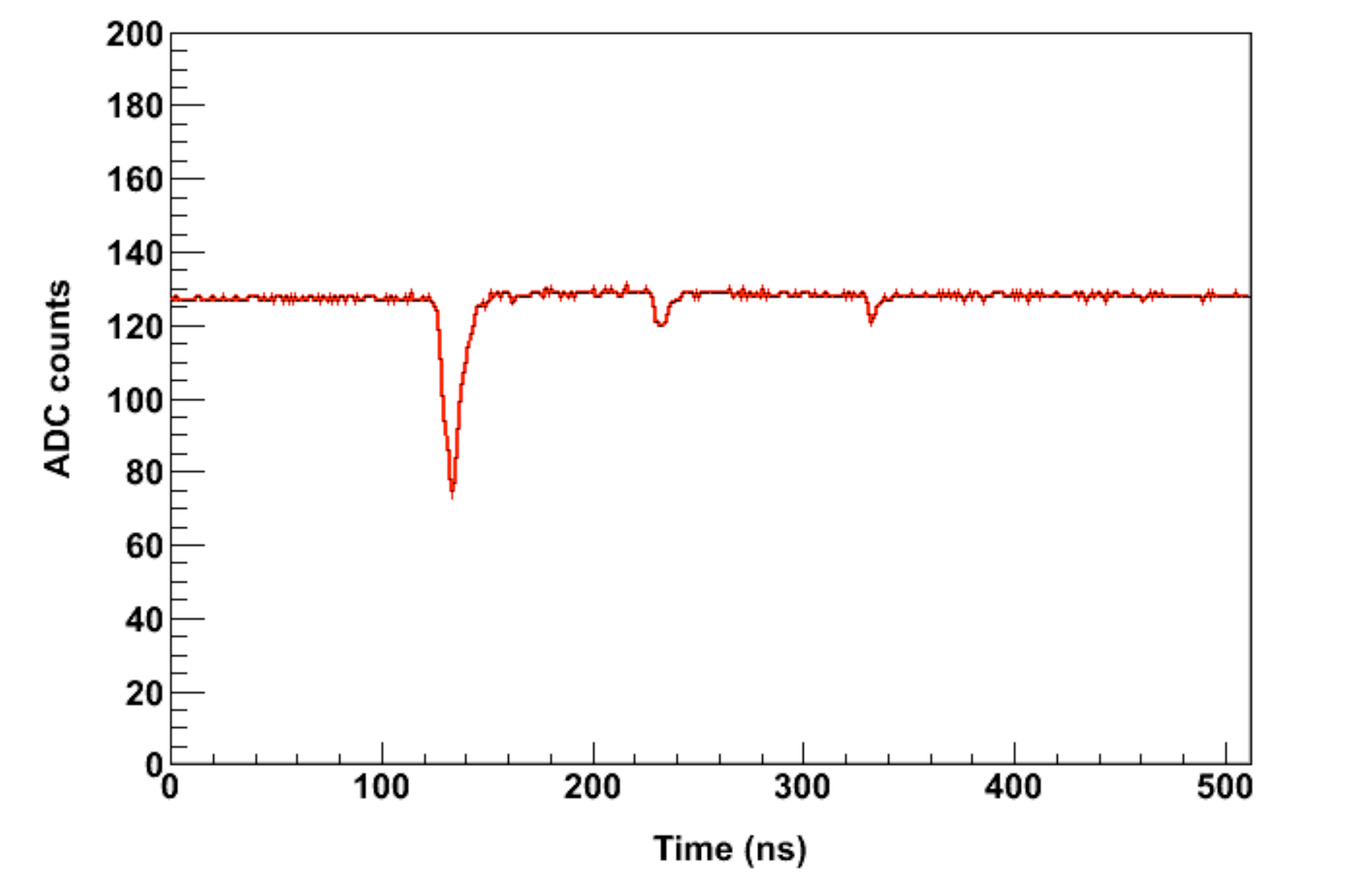}} 
{\includegraphics[width=2.in]{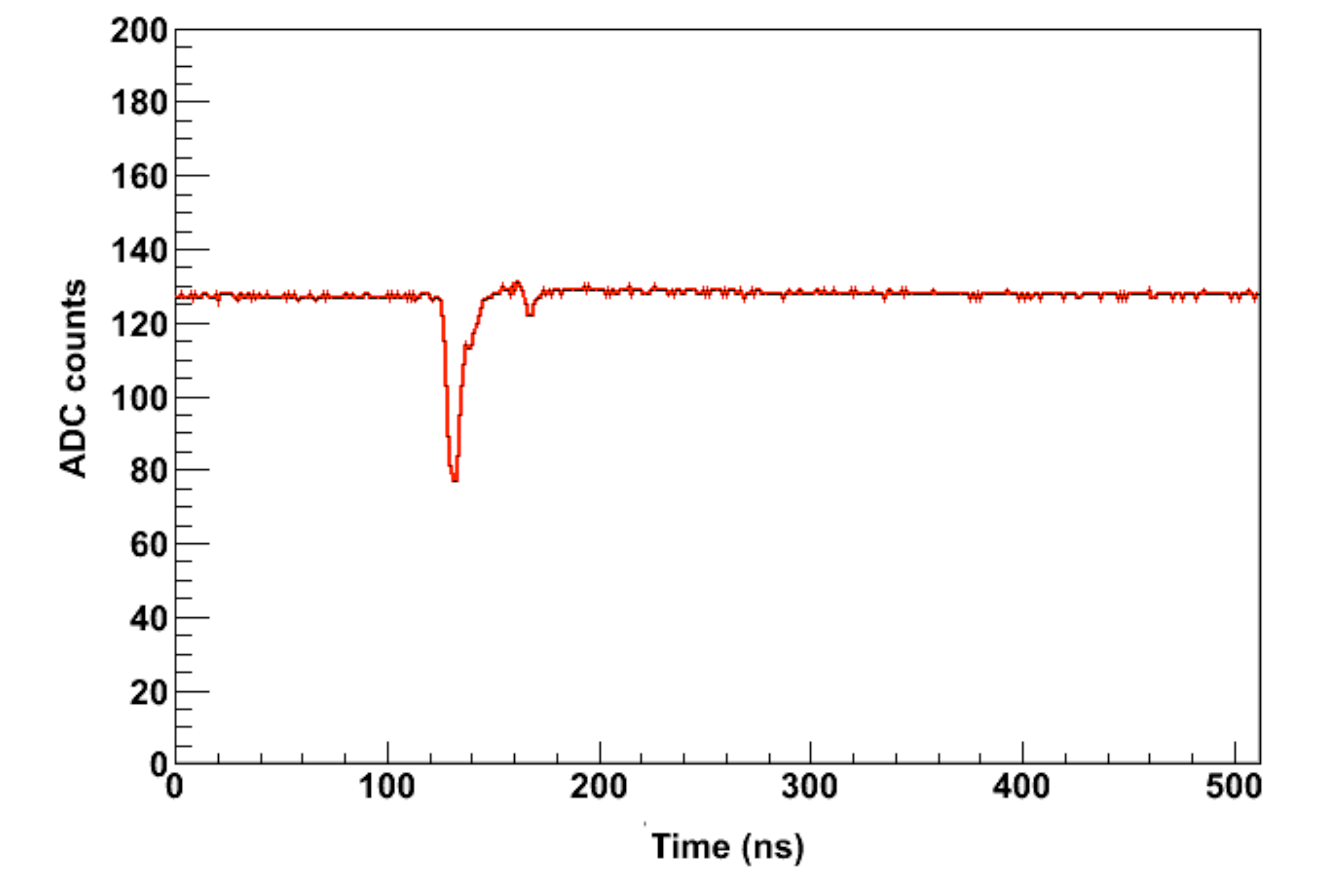}} 
{\includegraphics[width=2.in]{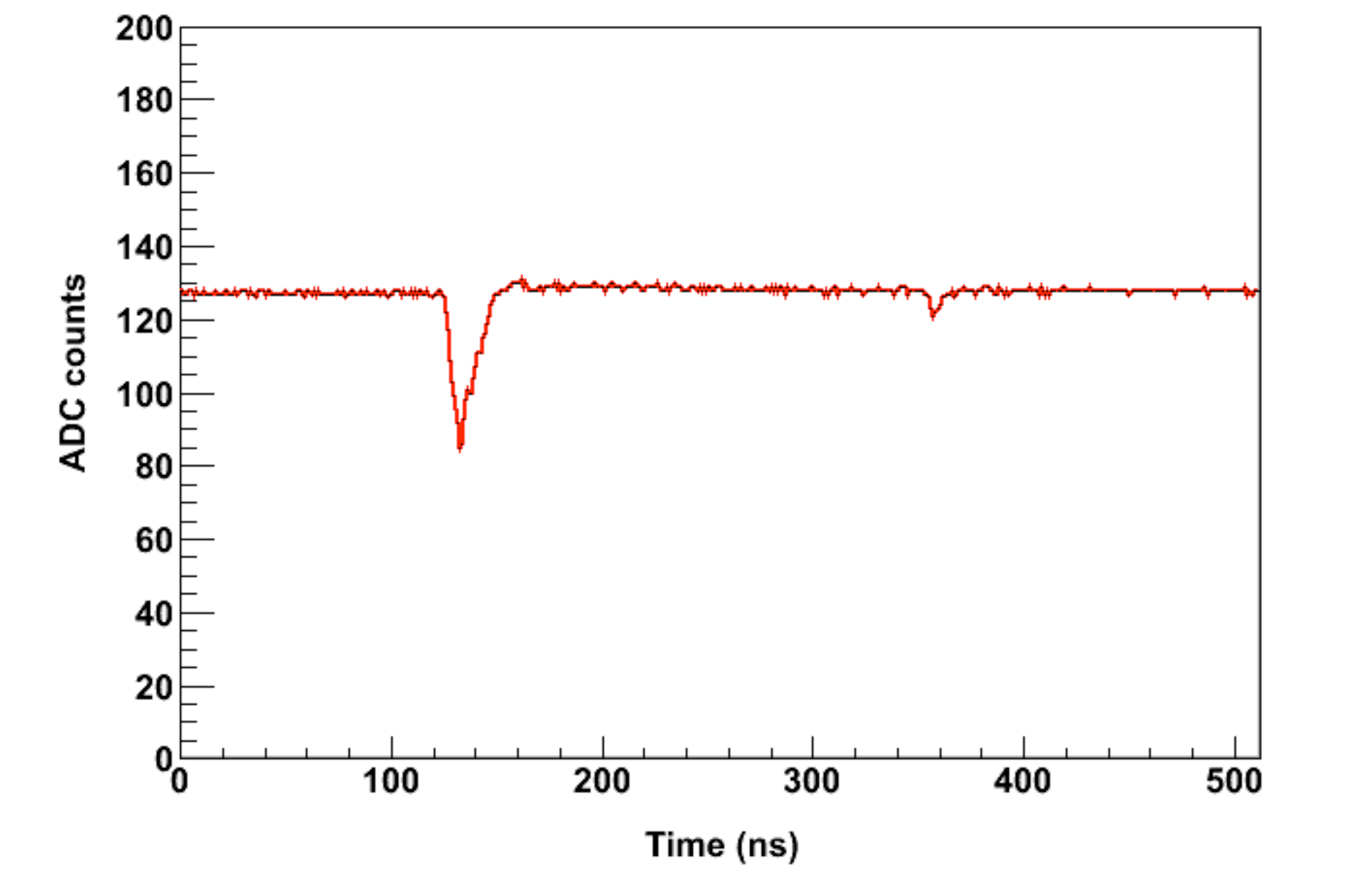}} 
{\includegraphics[width=2.in]{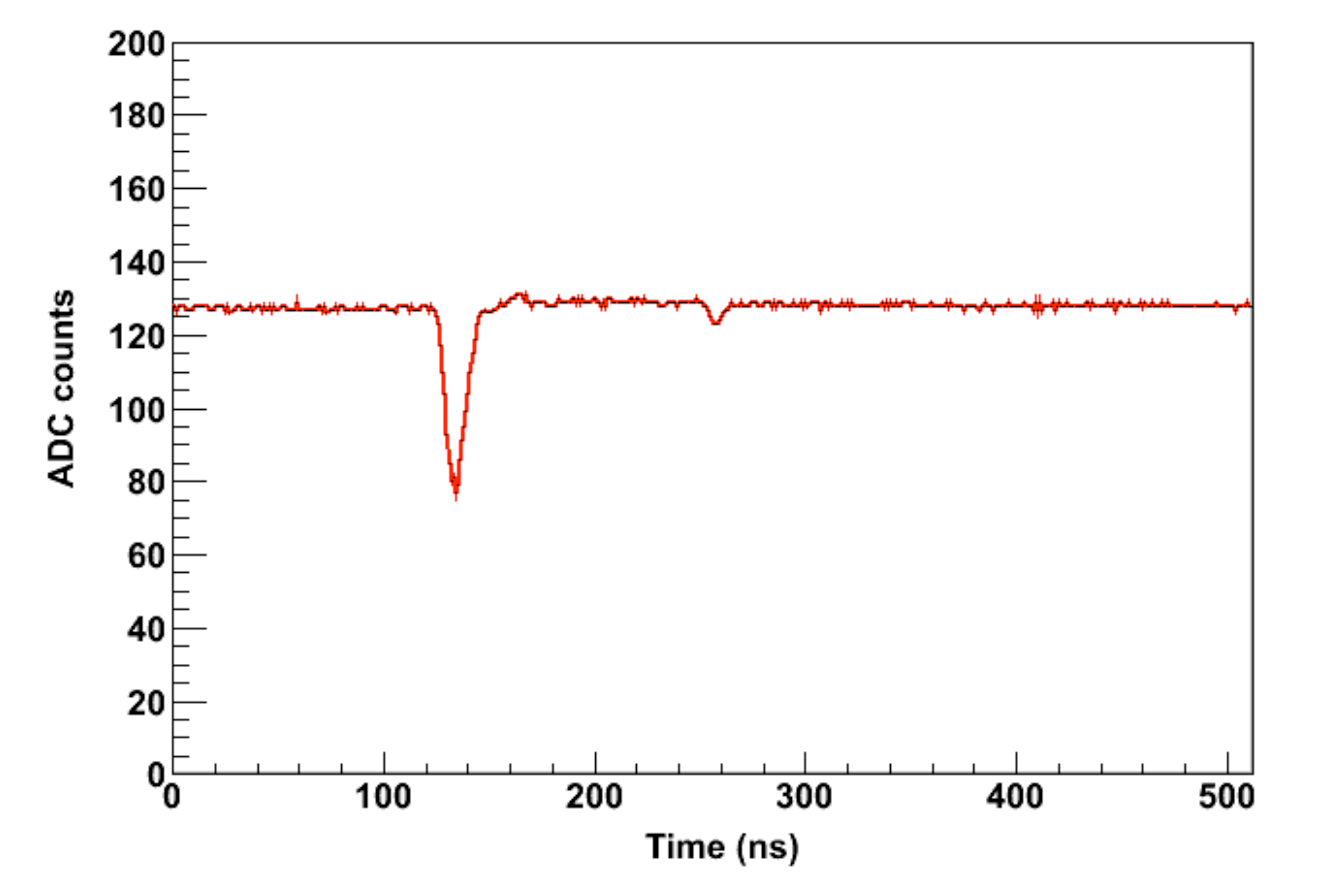}} 
\end{center}
\caption{Examples of waveforms after cuts described in Sec~\ref{cuts}.  The first pulse is early light 
which produces the trigger. The trigger level is set at a 15 ADC counts, 
which corresponds to just below 3 PE, as determined in Sec~\ref{analysis}.  
The early pulses (with time constant of $\sim 6$ ns) are produced by multiple UV
photons impinging on the bar.  Thus the early pulses vary in pulseheight.
Single-PE late light pulses can be also be seen in these events.
\label{typical} }
\end{figure}

Industrial grade liquid argon is used in this study. This has high
levels of impurities, which can vary by factors of two to three between
batches.  Two body collisions with impurities (for example Ar$_2^*$
+O$_2$ $\rightarrow$ 2Ar + O$_2$, and the analogous process with N$_2$)
reduce the late scintillation light which is emitted \cite{WARPoxy,
  WARPnit}.

Readout is performed using an Alazar Tech ATS9870 digitizer, run in
traditional mode with record headers enabled.  The channel is
configured to have an input range of $\pm$200 mV.  A trigger
is produced by a negative pulse with an amplitude that
exceeds 15 ADC counts, which corresponds to a peak voltage of $-23.44$
mV.  When a trigger is produced, 128 pre-trigger samples and 384
post-trigger samples are recorded at a sampling rate of 1 gigasample
per second, leading to a total recorded profile of width 0.512 $\mu$s
from both channels.  We store the pulse profiles to disk for later
analysis offline.

Figs.~\ref{typical} and \ref{cutevent} provides some illustrative
examples of event waveforms.  The first pulse, which is required to be
above the 15 ADC-count trigger, is consistent with the many photons
expected for the early light.  Despite the impurities in the
industrial grade argon, we do observe late-light pulses in the events.

\section{Conversion to Photoelectrons  \label{analysis}}

\subsection{Analysis Cuts for the Calibration Samples \label{cuts}}

\begin{figure}[t]\begin{center}
{\includegraphics[width=2.in]{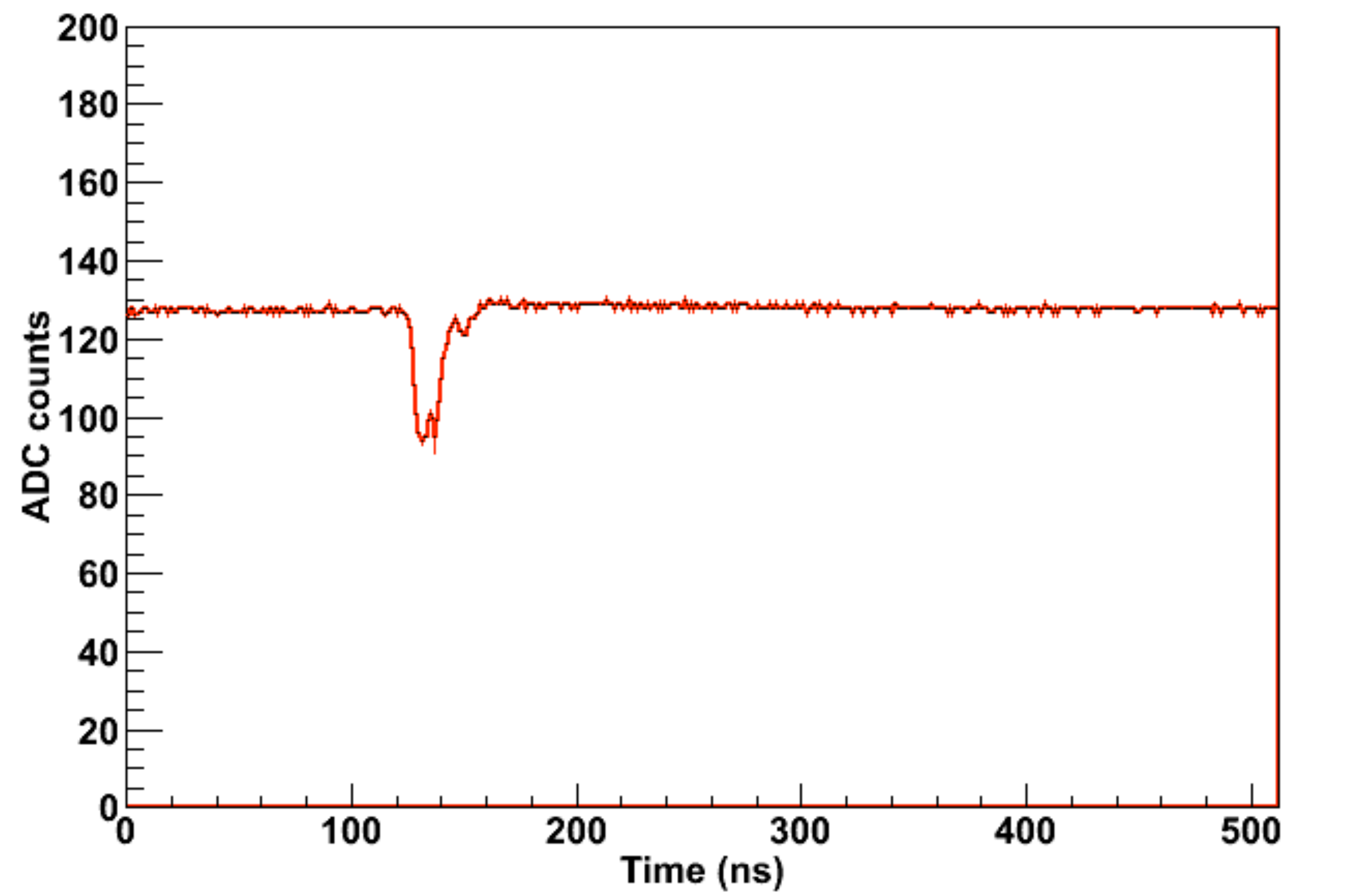}} 
{\includegraphics[width=2.in]{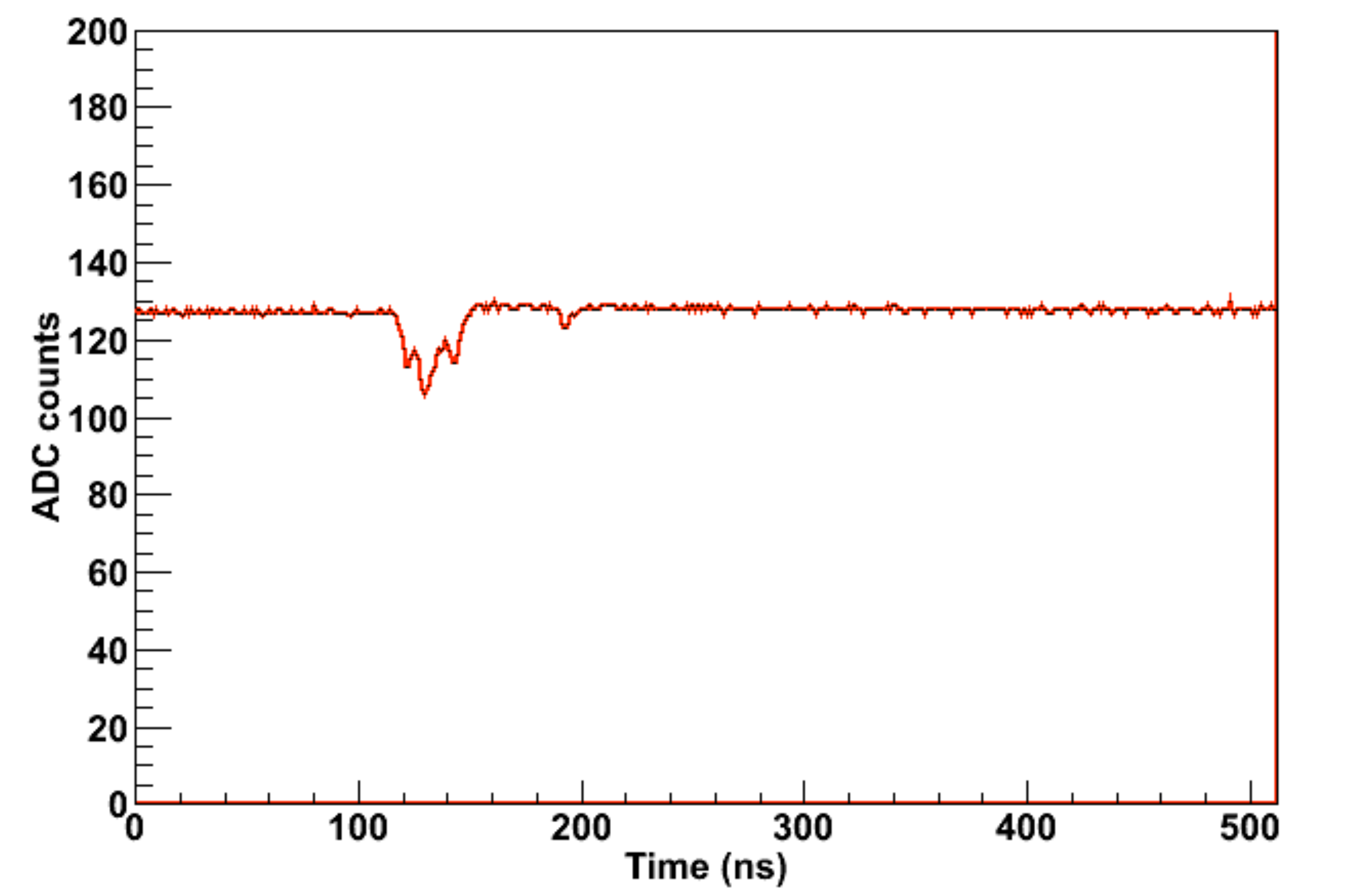}} 
\end{center}
\caption{Examples of waveforms removed by the cuts described in Sec.~\ref{cuts}.
\label{cutevent} }
\end{figure}

In order to perform the calibration, ``clean pulses,'' such as those
shown in Fig.~\ref{typical} are required.  However, the majority of
events have merged pulses, such are shown in Fig.~\ref{cutevent}.  We
apply cuts to obtain clean pulses as described in this section.
Because the cuts are strict, we use a high statistics run for this
study.  In this run,  data were taken with the source at 20 cm from the PMT
face.

The first analysis will use the late light.  We identify late pulses
as those that occur $>50$ ns after the trigger, to reduce overlap with
wide early signals, such as the case shown in Fig.~\ref{cutevent}
(left).  A late pulse is identified if the signal is 2 or more ADC
counts below the baseline, which cuts noise from the ADC that is at
the single count level. The result is a clean sample of low PE pulses
with no hardware trigger threshold, as is seen in Fig.~\ref{typical}.

The second analysis, which cross-checks the first, will use early
pulses.  To isolate early pulses where the signals arrived nearly
simultaneously, we place ``early pulse cuts'' on the events:
\begin{itemize}
\item the full-width-half-max is $<12$ ns;
\item the pulse is consistent with a single peak rather than multiple
  peaks.
\end{itemize}
The search for multiple peaks is initiated by a change of $+2$ counts
from the first minimum in ADC counts. Further peaks are identified if
there are either two consecutive changes of $-1$ ADC counts per
time-bin or one change of $-2$ or more ADC counts per time-bin.

These cuts result in the clean early pulses seen in
Fig.~\ref{typical}.  However, nearly 80\% of the events are removed by
the early pulse cuts.  As a result, using the high statistics sample
for this study is necessary.

\subsection{Calibration with Late Light}

\begin{figure}[t]\begin{center}
{\includegraphics[width=4.in]{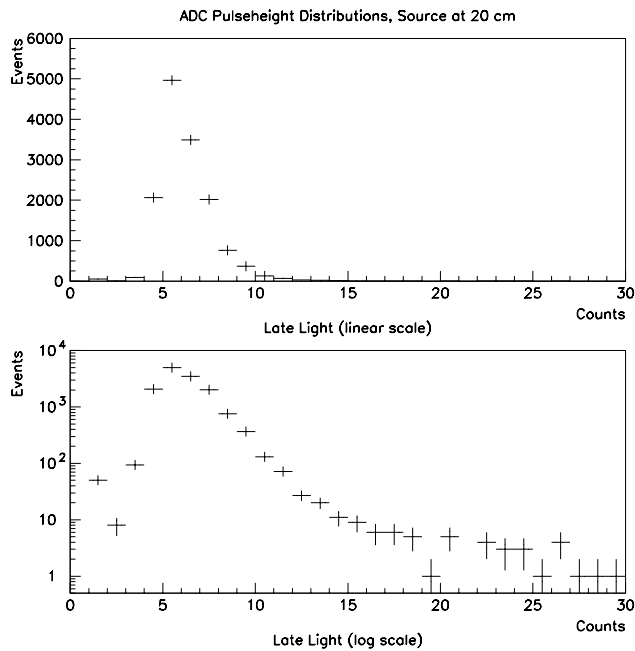}} 
\end{center}
\vspace{-1.25in}
\caption{Late light distributions from a high statistics measurement 
at 20 cm used to study the ADC-pulseheight-to-PE Conversion.  Late light
analysis cuts require bin 1 to be zero.
Top:  Pulseheights, in ADC counts, linear scale; Bottom: Same
plot log scale.  
\label{pepeak} }
\end{figure}

Late light can be used to calibrate the relationship between ADC
counts and 1 PE. The distribution of pulseheights for late light, as a
function of ADC count, is shown on a linear scale in Fig.~\ref{pepeak}
(top) and on a log scale on Fig.~\ref{pepeak} (bottom).  The first bin
is empty due to the analysis requirement which removes ADC noise.  The
events at two ADC counts, visible in the log plot, are consistent with
the tail of the ADC noise distribution, indicating that the ADC is
sensitive to low count events.  The distribution is asymmetric with a
peak at 5.7 ADC counts.

\subsection{Calibration with Early Light \label{early}}

\begin{figure}[t]\begin{center}
{\includegraphics[width=4.in]{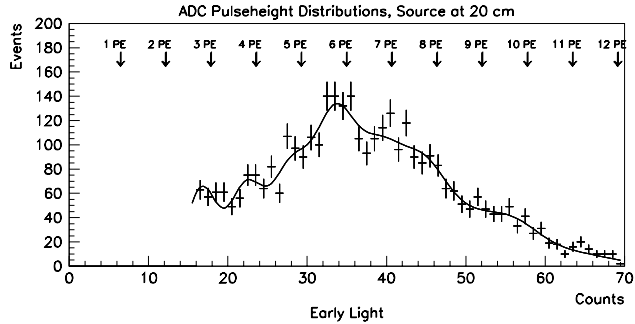}} 
\end{center}
\vspace{-2.75in}
\caption{Early light distributions from a high statistics measurement 
at 20 cm used to study the ADC-pulseheight-to-PE Conversion. 
Distribution is shown in ADC counts, with cuts described in the text.   Fit is to ten equally spaced Gaussians, as discussed in the text,
corresponding to PE peaks 3 to 12.   Arrows indicate the peak position based on the fit.
\label{peearly} }
\end{figure}

Fig.~\ref{peearly} shows the distribution of early light
pulseheights in ADC counts.  The distribution appears to have peaks,
which are seen in all pulseheight measurements when cuts are applied.
This would arise if the data represent a collection of Poisson
distributions, each centered at an integer number of PE.  To test this
model, the data are fit to Gaussians, as an approximation to Poissons,
with means required to be at integral multiples of the same
parameter. The resulting fit gives a conversion of 5.7 ADC counts per
PE, which is in agreement with the calibration from the late light
peak. The arrows in Fig.~\ref{peearly} indicate the fit
position of each PE peak. In the fit, the Gaussian widths are required
to vary according to the square root of the number of PE, multiplied
by a common overall parameter, found to be 1.4 ADC counts.  The
$\chi^2$/DOF is 1.4, which is good given the simplicity of the model.

\subsection{Final Calibration:  Pulseheights to PE}

The late and early light analyses are consistent with 5.7 ADC counts
per PE. Therefore, in the following discussion, the pulseheights are
rebinned such that ADC counts 3 to 8 correspond to 1 PE, 9 to 14
correspond to 2 PE, and so forth.

\section{Calculation for Ideal Performance \label{ideal}}

\bigskip%
\begin{table}[tbp] \centering
{\footnotesize
\begin{tabular}
[c]{|l|l|c|c|c|}\hline
\multicolumn{5}{|c|}{Parameters Used in Ideal Calculations} \\ \hline
\#  & Parameter & Value in Calc. & Source & See Sec(s)\\ \hline 
\multicolumn{5}{|c|}{Related to UV light production} \\ \hline
1 & Early ($<10$ ns) UV $\gamma$/MeV (MIP) & 7600 & Ref. \cite{Doke} &
\ref{ideal} \& \ref{lar20}\\ 
2 & Light reduction factor for $\alpha$  & 0.72 & Ref. \cite{Doke} & \ref{ideal}\\
3 & Light reduction factor for $p$  & 0.81 & Ref. \cite{pquench} & \ref{lar20}\\ \hline
\multicolumn{5}{|c|}{Related to Geometry of Teststand} \\ \hline
4& Acceptance of UV light & 0.33 & calculated & \ref{ideal} \\ \hline
\multicolumn{5}{|c|}{Conversion and Capture in the Bars} \\ \hline
5 & UV/visible $\gamma$s, evaporative coat & 1.0 & Ref.~\cite{VicPrivate} & \ref{ideal} \& \ref{lar20}\\ 
6 & Response, bar coating to evap. & 0.1 & measured &  \ref{ideal} \& \ref{lar20}\\ 
7 & Capture fraction & 0.05 &  calculated &  \ref{ideal} \& \ref{lar20}\\  \hline
\multicolumn{5}{|c|}{PMT response} \\ \hline
8 & QE of 7725 PMT & 0.25 & Ref.~\cite{Hamamatsu} &  \ref{ideal} \& \ref{lar20}\\ 
9 & Cryogenic modification factor & 0.8 & Ref.~\cite{Meyer} &  \ref{ideal} \& \ref{lar20}\\ \hline \hline
\multicolumn{5}{|c|}{Combining Parameters to Calculate Efficiencies} \\ \hline  & Efficiency & Value & Combined  & See Sec(s)\\ 
 &  &  & Params. & \\ \hline \hline
10 & Efficiency to convert and capture & 0.005 & $5\times 6 \times 7$ &  \ref{ideal} \& \ref{lar20}\\
 & Total Ideal Efficiency & 0.001 & $8 \times 9 \times 10$  & \ref{ideal} \& \ref{lar20}\\
\hline
\end{tabular}}
\caption{Parameters used in ideal calculations presented in Secs.~\ref{ideal} and \ref{lar20} for light production and bar acceptance.  \# is the parameter number appearing in the text and in the last two rows of this table, under
``Combined Params.'' ``MIP'' means minimum ionizing particle. ``QE'' is quantum efficiency. ``Measured'' (``calculated'') indicates that this is a measurement (calculation)  by the authors described in the text.   \label{calcparam}}%
\end{table}%

Our goal is to compare the actual performance of the lightguides to
predicted performance.  The performance of the lightguide depends upon
the response of the bar to the UV light, the efficiency for capturing
the visible light, the efficiency for guiding the light to a PMT, and
the efficiency for the PMT response.  The ideal bar would have 100\%
efficiency for guiding the light.  In this section we consider the
other contributions, in order to calculate the ideal performance of
the system in the absence of attenuation.  The parameters used in the
calculations in this section are summarized in
Tab.~\ref{calcparam}. The prediction of the average PE proceeds in two
parts: 1) the calculation of the UV light per $\alpha$ which will
reach the guide; 2) the calculation of the light which will capture
and produce photoelectrons.

In order to calculate the UV photons impinging on the guide, we assume that each 5.3 MeV $\alpha$ exits the source without
energy loss.  In the case of liquid argon, there is
a suppression of scintillation light for a heavily ionizing particle
compared to a minimum ionizing particle (MIP) \cite{Hitachi}, which is 
$(72 \pm 4)\%$ \cite{Doke}. A MIP produces 7600 UV photons per MeV
in the first 10 ns \cite{Doke}, the time-frame most useful for triggering
and vetoing.  Combining this information, $3\times 10^4$ prompt photons
are produced.

The UV light is emitted isotropically in the LAr.  Because the
$\alpha$ travels a negligible distance in the LAr, the light is
produced within the well, just outside of the silver foil.  The walls
and back of the well of the disk obstruct the light, limiting the
acceptance for light to hit the bar.  Based on the geometry of the
well, we find that 33\% of the light should be within the
acceptance of the bar, with an estimated 20\% systematic error
due to the irregular shape of the source.
Based on the above, we find $1\times10^4$ UV photons hit the
bar, per $\alpha$ emitted into the LAr
(parameters $1\times2\times4$ in Tab.~\ref{calcparam}, multiplied by
5.3 MeV).

Next we must calculate the efficiency of the bar to shift UV light and
capture visible light.  We will assume that the surface of the bar is
fully covered with TPB-polystyrene coating.  We begin with the
efficiency for pure TPB to convert 128 nm UV to visible light.  This
was measured by studying evaporative coatings by
Ref. \cite{VicPrivate}.  Next, we must account for the loss of light
due to the additional polystyrene in our coating.  We have measured
the relative response of coatings to evaporative coating using a
vacuum spectrometer.  The reduction factor is $\sim 10\%$ with a
spread of about 25\% in the measurements.  Once the visible light is
emitted, only a small fraction will be captured.  The relative indices
of refraction lead to 5\% capture efficiency in the bar.  The total
efficiency of the bar, in the case of infinite attenuation length, is
0.5\% (parameters $5\times 6 \times 7$ on Tab.~\ref{calcparam}).

Finally, we also must consider the quantum efficiency (QE) of the PMT.  The
7725 tube listed in the Hamamatsu catalogue \cite{Hamamatsu}, which is not cryogenic,
has a 25\% QE at the wavelengths of TPB emission.  Modified tubes for cryogenic running have been shown to have
lower QE \cite{Meyer}.  We will assume a QE of $20 \pm 2\%$ for the 
7725-mod PMT.

The resulting total efficiency of the system is obtained by
multiplying the efficiency of the bar with the QE of the PMT.  The
result is an efficiency of 0.1\%.  Multiplying by the predicted UV
photons impinging on the bar, one predicts an average of 10 PE with a
systematic error of about 33\%.  This is the predicted average PE for
a bar, with a perfectly smooth, full-coverage TPB-polystyrene
coating.

\section{Performance Measurements}

In the real system, the light which is detected is most affected by
the quality of the coating.  Small uncoated regions will reduce the
number of visible photons produced per UV photon impinging on the bar.
Roughness in the coating will cause loss of visible light as it is
guided to the PMT.  We try to separate these issues through two
measurements.  First, we measure the the light production at a
position which is as near to the PMT as possible.  The average of this
distribution can be directly compared to the calculated ideal average.
Differences would arise from small uncoated regions on the surface 
of the bar.
Next, we make measurements with the source placed at various positions
along the bar.  In principle, this would allow a measurement of the
attenuation along the bar.  In practice, this result is so complicated
by local variations in the coating coverage at each position that it
is difficult to achieve a good attenuation length measurement.

\subsection{Test of Coating Variation at 10 cm \label{pe}}

\begin{figure}[t]\begin{center}
{\vspace{-0.5in}\includegraphics[width=4.in]{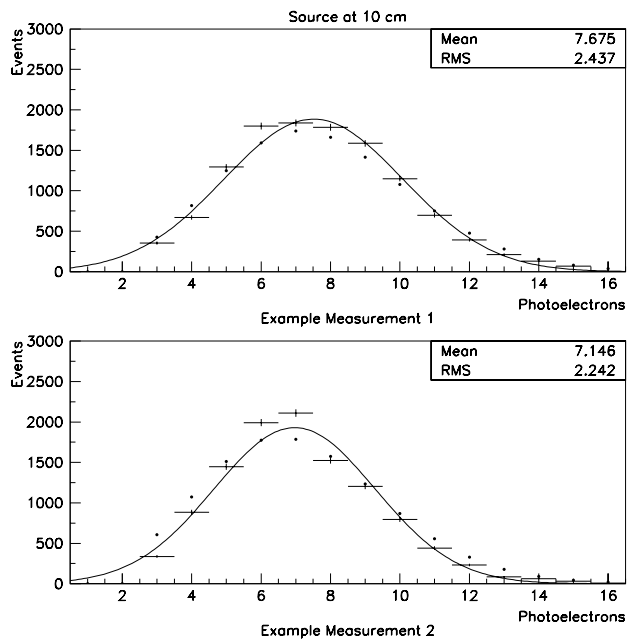}} 
\end{center}
\vspace{-1.in}
\caption{Two examples of PE distributions, measured at 10 cm, with no
cuts applied.  The mean and
sigma of each data distribution appears on the plot.  The difference in position of the two peaks may be due to variations in the coating in the local area; PE measurements are compared to the Gaussian fit and the Poisson prediction (dots).
\label{tenpeak} }
\end{figure}

For an ideal light guide, with no attenuation, the PE distribution is
predicted to have a mean at 10 PE.  To test the actual response, we
must use a source placed at 10 cm from the PMT, due to the design of
the test stand.  Assuming long attenuation lengths, the result should
be comparable to the ideal.  We will revisit this assumption in the
discussion of attenuation length (Sec.~\ref{loss}).

In this test, we repeat the following process: the source is placed
onto the rod, data are taken, and the source is then removed. Each
time,  the source is replaced to within $\pm 0.25$ cm of the vertical
position.  The lateral position is chosen
randomly, but near to the center of the bar.  This allows us to study 
variations in the coating in the local area of the measurement. 

The observed variations were large.  The peaks of the PE
distributions, which are presented without cuts, varied between 7 and 8
PE.  Fig.~\ref{tenpeak} shows two examples.  The variations most likely
indicate that some regions of the surface are more fully covered with
TPB-polystyrene coating than other areas.  This can be further
investigated in the future, as our HVLP coating techniques improve.

While the peaks of the distributions vary, qualitatively, the shapes
are consistent.  This can be seen in the two examples of
Fig.~\ref{tenpeak}.  The distributions are asymmetric, favoring lower
PE, as is emphasized when compared with Gaussian curves.  In this
sense, the distributions are Poisson-like.  However, they are
consistently more peaked than true Poissons, which are shown by the
closed circles.  This may be indicative of a systematic effect
in our conversion to PE.

With the simple assumption of negligible attenuation of light over 10
cm, the performance is between 7 and 8 PE. The response is
consistently less than ideal, but high enough to be useful, as we will
show in Sec.~\ref{lar20}.  In the next section, we show that the low
measured response may simply be a matter of extrapolation to zero
position along the bar.  However, the variations in the peak positions
already point to gaps in the coating coverage.  Less than 100\%
coverage would will reduce the overall response.  Improving the
surface coverage is the next goal in this R\&D program.

\subsection{Attenuation Along the Bar \label{loss}}

Detected light from points along the bar is a combination of light
produced at the source and light lost during transport along the bar.  In the
previous section, we showed that light production has large
variations.  Separating this from effect from light loss along the bar is very
difficult.
Therefore, in this section we extract only qualitative information
from measurements with the present design.

We measure the attenuation length by comparing trigger rates as the
source is moved along the bar.   Rates are background subtracted, but
no other cuts are applied.  Backgrounds, which are primarily cosmic
rays, are measured with no-source runs.    

Multiple measurements were made at 10, 20, 40 and 50 cm. A scan was
made in the region between 20 and 22 cm.  Each time a measurement is
made, the source is removed and re-attached.    

\begin{figure}[t]\begin{center}
{\includegraphics[width=4.4in]{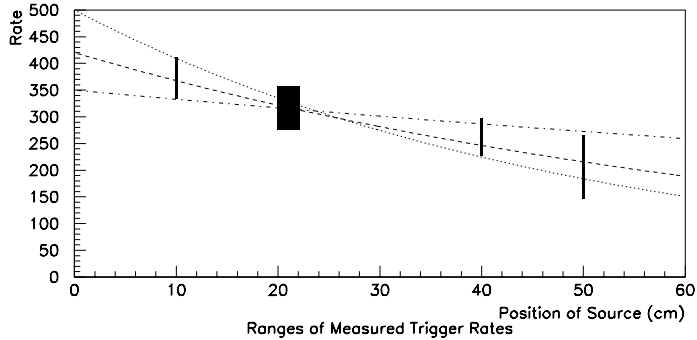}} 
\end{center}
\vspace{-3.in}
\caption{Ranges of rate measurements as a function of source position 
along the bar.   Multiple measurements were made at 10, 20, 40 and 50 cm,
and then through the 20 to 22 cm region.  The exponentials indicates 
attenuation length of 50 cm (dotted), 75 cm (dashed) and 200 cm (dot-dashed).
\label{atten} }
\end{figure}

At any given position for the source on the bar, multiple measurements
resulted in a range of rates, as was expected given the results of
Sec.~\ref{pe}.  In Fig.~\ref{atten} we use the height of the boxes to
indicate the range of measured rates at each location.  The variations
are on the order of 10\% at each point, which is consistent with the
variation in peak position presented in Sec.~\ref{pe}. 

As a result of these variations, we can only obtain a qualitative
sense of the attenuation length.  On the assumption that the
attenuation length can be described by a single exponential,
Fig.~\ref{atten} shows curves for 50, 75 and 200 cm attenuation
length.  As a conservative estimate, the attenuation length is $>$50
cm in this simple model.   

Revisiting the assumption in Sec~\ref{pe} of negligible attenuation
over 10 cm, the extrapolation to 0 cm may indicate a correction
between zero and $\sim$20\%.  However, we note that of the three
examples shown, the extrapolation that is most consistent with the measured rate
from the source described in Sec.~\ref{test} is the 200 cm attenuation
length.

\section{Discussion on Use in LArTPCs \label{lar20}}

\begin{figure}[t]\begin{center}
{\includegraphics[width=3.5in]{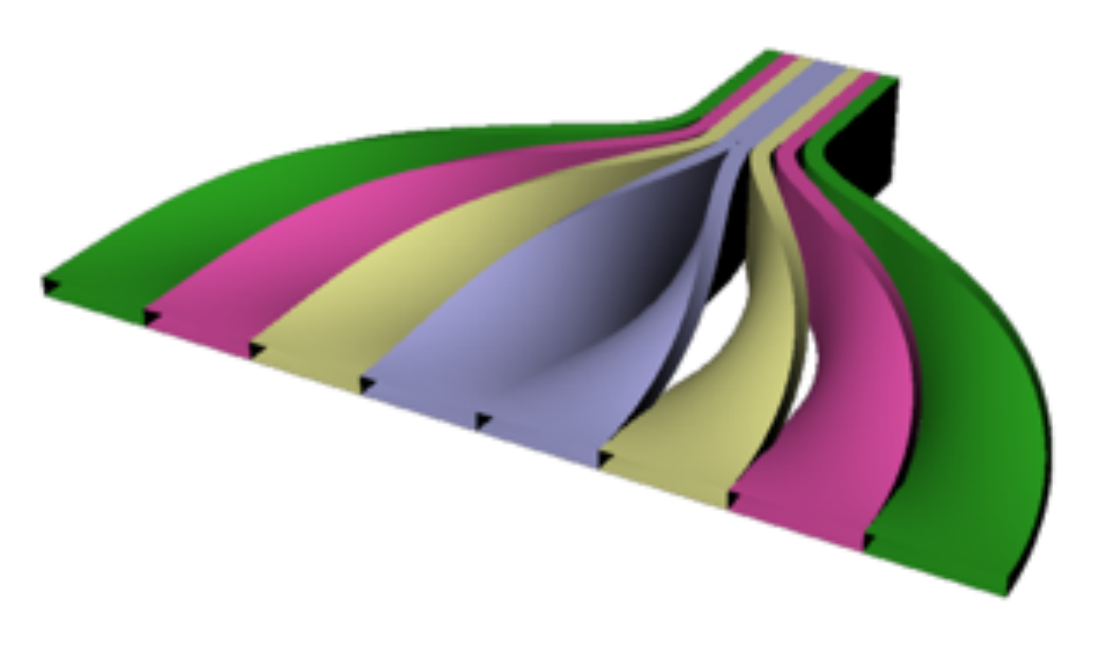}} 
\end{center}
\caption{Illustration of multiple bars bent to adiabatically guide light to a single PMT.
  \label{twisted} }
\end{figure}

One could instrument a large LArTPC detector with a trigger system
constructed from the acrylic bars.   Here, we consider the prospects
for building such a system today, given the capability we present in
this paper.  This sets a minimum for what can be accomplished with
such a system, since the bars can surely be improved in the future.

The basic detector element will be a paddle constructed of eight 2.54
cm width bars with a 1 m long active area.  At the top, the uncoated
portion of the acrylic will be bent to form an adiabatic path for
light to the PMT (see Fig.~\ref{twisted}).  If 8 bars come to one 2
inch PMT, then the collection area of this paddle is 2032 cm$^2$.  We
consider collection needed for a volume of 250 cm $\times$ 250 cm
$\times$ 1000 cm, and a surface area of $1.1\times10^6$ cm$^2$, which
is the size of the MicroBooNE Experiment \cite{MicroBooNE}.

Based on the
measurements presented in Secs.~\ref{pe} and \ref{loss}, we assume
that by choosing bars with highest quality coating, we can achieve 8
PE on average and with a 2 m attenuation length.  This results in 80\%
of the ideal expectation and an average transmission along the bar of
77\%, hence an overall reduction factor of 60\%.  While these
assumptions are slightly aggressive, the bars are very inexpensive to
produce and so a plan to choose only bars with the best response is
feasible.

For the sake of discussion, we will assume the trigger will be based
on summed pulses passing an ADC theshold.  We determine the paddle
coverage required to trigger on particles with $>99\%$ efficiency,
which requires a threshold of less than 1 PE and at least 5 PE
detected on average.  We show that for 40 MeV protons, this is
straightforward to achieve today.  For the case of 5 MeV electrons, a
factor of two improvement in the paddle response is desirable.

\subsection{Example 1: Efficient triggering on a 40 MeV proton}

Low energy neutral current elastic (NCEL) scattering events which
produce a proton with 40 MeV of kinetic energy are used as the
benchmark for light collection in the MicroBooNE experiment
\cite{MicroBooNE}. To examine the feasibility of meeting this
benchmark with paddles, we first calculate the coverage required if we
had ideal paddles, using the parameters in Tab.~\ref{calcparam}. We
then adjust for the real response reported in Secs.~\ref{pe} and
\ref{loss}.  Lastly, we correct for factors related to running within
a TPC, including suppression due to the electric field.

Protons produce 81\% less scintillation light than a MIP particle
(parameter 3 in Tab.~\ref{calcparam})
\cite{pquench}.  To obtain the total UV photons produced by a
40 MeV proton, multiply parameters $1 \times 3$ in
Tab.~\ref{calcparam} by 40 MeV to get $2.5 \times 10^5$ UV photons.

Next, we convert the number of UV photons to detected PE in an ideal
paddle.  Begin by considering a detector with 100\% coverage by ideal
paddles, which have a total efficiency of 0.1\% (last row of
Tab.~\ref{calcparam}).  In this case, 250 PE would be detected.  No
LArTPC can have 100\% coverage because materials of the HV cage and
TPC intercept light.  Also, to save cost and effort, one would only
install sufficient coverage to guarantee efficient triggering.  In the
case of the ideal paddles, this means that only 2\% coverage is
required to detect 5 PE on average.  This corresponds to 11
paddles.

As discussed above, we assume that we can select bars which deliver 8
PE on average and have a 2 m attenuation length.  Combining these two
effects results in a reduction factor of 60\% compared to the ideal
case.  As a result, 18 paddles will be required to collect 5 PE per 40
MeV proton on average.
 
The application of an electric field, which is necessary for charge
collection in an LArTPC, suppresses scintillation light.  From Fig. 2
of Ref.~\cite{Kubota}, the suppression factor will be about 66\%.  As a
result, 27 paddles are required.

This calculation omitted some factors specific to the design of a
given LArTPC.  For example, the paddles would most conveniently be
placed behind the TPC plane, where there is no electric field.
In this location TPC wires intercept the light.  This effect
depends on the wire spacing and diameter, and the number of wire
planes. In MicroBooNE, about 80\% of the light is transmitted through
the wires.  A second design-specific issue is reflections of the UV
light, which depends upon the materials in the LArTPC.  ICARUS found
that reflections increased the observed light by about 20\%
\cite{ICARUS}.  If we assume that
design-specific effects roughly cancel, then about 30 paddles are
required to assure efficient triggering on a 40 MeV proton.

\subsection{Example 2: Efficient triggering on a 5 MeV Electrons}

Efficient triggering on 5 MeV electrons produced by supernova burst
events or solar neutrino interactions may be required in a future
LArTPC.  At 5 MeV, the electron is a MIP and so $3.8\times10^{4}$ UV
photons are expected.  In the ideal case, for 100\% coverage, 38
PE will be observed, and 13\% coverage is required to assure 5
PE on average.  Applying the correction factors for
measured-to-ideal response and attenuation length yields 28\%
coverage.  Once the factor of 66\% is applied for loss of light due to
the electric field, 42\% coverage is required with today's lightguide
technology.

It is most convenient to place paddles behind the TPC plane, which
represents 22\% of the surface area. If we restrict installation to
only this space, then we require $\times 2$ better response from the
paddles.  This may be achieved by a number of improvements.  The first
is to produce paddles which meet the ideal calculation and have
negligible attenuation losses.  This ideal calculation was for a
3-to-1 polystyrene to TPB mixture, by weight.  Improving this ratio,
while avoiding crystallization, would increase the emitted light per
UV photon.  Also, we must investigate PMTs with increased quantum
efficiency \cite{Mirzoyan}, and other efficient light collection
technologies, such as silicon photomultipliers \cite{Spooner}.

\section{Conclusions}

In conclusion, lightguides can provide a cost effective alternative to
the present light collection system in LArTPCs.  This paper has
demonstrated first light collection using acrylic lightguides in
argon.  We have begun measuring and optimizing the relevant
parameters.  Improvement in the quality of the coating coverage and
the smoothness of the coating have been identified as important issues
to be addressed with future R\&D.  Nevertheless, we have shown that
with the present performance, paddles made from these bars are
efficient for triggering on 40 MeV NCEL events.

\section*{Acknowledgments}

The authors thank the Guggenheim Foundation and the National Science
Foundation for support.  We thank Lindley Winslow for contributions to
the DAQ code.  We appreciated input from Sten Hansen and Stephen
Pordes, of Fermilab.  We thank Charlie Abbott for coating the bars. We
thank the members of the MicroBooNE PMT Group, Bruce Baller and Stuart
Mufson for valuable comments on this paper.

\bibliographystyle{unsrt}

\bibliography{lightguides_bib}

\end{document}